# Unveiling potential candidates for rare-earth-free permanent magnet and magnetocaloric effect applications: a high throughput screening in Fe-N alloys

Qiang Gao [a,b,&,*], Ergen Bao [b,&], Ijaz Shahid [b], Hui Ma [b], Xing-Qiu Chen [b,*]

a *School of Science, Shenyang University of Technology, Shenyang 110870, China*

b *Shenyang National Laboratory for Materials Science, Institute of Metal Research, Chinese Academy of Sciences, Shenyang, 110016, China*

**Abstract**:

Based on high-throughput density functional theory calculations, we have found 49 ferromagnetic cases in $Fe_xN_{1-x}$ ($0<x<1$) compounds, focusing especially on permanent magnet and giant magneto-caloric effect applications. It is found that 15 compounds are potential permanent magnets with a magneto-crystalline anisotropy energy more than 1 MJ/m$^3$, filling in the gap of application spectrum between high-performance and widely used permanents. Among the potential permanent magnets, $Fe_2N$ can be classified as a hard magnet while the other 14 compounds can be classified as semi-hard magnets. According to the calculations of magnetic deformation proxy, 40 compounds are identified as potential giant magnetocaloric effect candidates. We suspect that Fe-N compounds provide fine opportunities for applications in both rare-earth free permanent magnets and magneto-caloric effect.

## 1. Introduction

Nowadays, the advanced information and green energy technologies are becoming much more important than ever before for providing us efficient and convenient life [1,2], with the fast-expanding requirement of hybrid-electric vehicles, robotics, wind turbines, and automation. The consumption of energy is becoming larger and larger, leading serious greenhouse gas emissions and chemical pollution to the environment [1,3,4]. In the field of information technology, more permanent magnets are required to design devices with desired rate of transformation, storage, and transmission speed [2,5,6]. The permanent magnet is the fundamental basis to realize environmentally friendly and high-performance energy and information devices. In applications, the excellent permanent magnet is eager to have a sizable coercivity, a remarkable saturation magnetization ($M_S$), and a significant high Curie temperature (at least above room temperature). The upper limit of the macroscopic coercivity is intrinsically determined by the microscopic magneto-crystalline anisotropy energy (MAE), which is associated with spin-orbit coupling (SOC) effect. At present, the commercially large scale applied permanent magnets are the rare-earth based compounds Sm-Co (MAE=17 MJ/m$^3$, $M_S$=910 kA/m) and Nd-Fe-B (MAE=5 MJ/m$^3$, $M_S$=720 kA/m) with a large energy density

[&] *These authors contributed equally to this work and should be considered co-first authors.*

[*] *Corresponding authors.*

*E-mail address*: waveflying@163.com, lxygaoqiang@sut.edu.cn (Q. Gao); xingqiu.chen@imr.ac.cn (X.-Q. Chen).

$(BH)_{max}$ [7]. Notably, the rare earth element is actually very rare and expensive, where the pollution problem also occurs during the production process. In commerce, the other widely used permanent magnets are transition metal-based compounds AlNiCo (MAE: 0.04 MJ/m$^3$, $M_S$: 50 kA/m) and ferrites (MAE: 0.03 MJ/m$^3$, $M_S$: 125 kA/m) with the relatively small energy density but cheaper cost [8]. Apparently, there is a remarkable gap in the application spectrum between such two types of commercial permanent magnets, waiting for the design of more permanent magnets with lower cost and reduced pollution. On the other hand, the realization of giant magnetocaloric effect (MCE) in magnets $Gd_5Si_2Ge_2$ [9] and $LaFeSi_{13}$ [10] has accelerated the studies and applications of high-efficiency refrigeration, playing an important role in green energy technology with environmentally friendly production process [11]. In this regard, it is a good chance to design rare-earth-free permanent magnets and giant magnetocaloric effect candidates with less cost to fill in the empty region of the application spectrum for advanced information and green energy devices [7,12].

The pristine α-Fe has a saturation magnetization as large as 217.6 emu/g (1710.34 emu/cm$^3$) [13] at room temperature but suffers from a significantly weak magneto-crystalline anisotropy energy of 1.34 μeV /Fe (0.036 MJ/m$^3$) [14] due to the underlying cubic symmetry. To design rare-earth-free gap permanent magnets, the tetragonal distortion should be induced into the cubic Fe by means of alloy technology, substitutional and interstitial doping as well as imposed strain [15,16]. For example, DFT calculations predict that by alloying, the Fe-Co system achieves various tetragonal distortions with an enhanced MAE in the order of 700-800 eV/at. and a comparable magnetization (≈1.6~2.3$\mu_B$/at.) to the pristine α-Fe [17], which is validated by the following experiments by observing a MAE of 2.9 MJ/m$^3$ with a magnetization of 2.5 $\mu_B$/at. at 400 K in the $Fe_{0.36}Co_{0.64}$/Pt superlattices [18]. However, the tetragonal distortion of Fe-Co is relaxed when the thickness is larger than 2 nm, hindering the application [19]. Moreover, both theoretical and experimental studies have revealed that the tetragonal distortion in Fe-Co thin film can be stabilized by inducing interstitial dopants of carbon [20] and boron [21], where the MAE is as large as 0.5 MJ/m$^3$ with a doped ratio of 4 at% for B [21]. Following this idea, the interstitial dopants (B, H, C, N) have been induced to cubic full Heusler compounds to design rare-earth-free gap permanent magnets by systematic high-throughput screening calculations, where 32 promising candidates fill in the application spectrum gap with a substantial out-off-plane uniaxial MAE (larger than 0.4 MJ/m$^3$) [22]. High-throughput screening for MAB phase compounds predicts 23 rare-earth-free permanent magnets with a uniaxial MAE larger than 0.4 MJ/m$^3$ and 99 promising MCE candidates with a magnetic deformation $\sum_M$>1.5% [23]. Using RF magnetron sputtering technology, the $\alpha''$-$Fe_8N$ thin film is grown on the substrate of MgO [001] [24]. The saturation magnetization is observed to be about 1750 emu/cm$^3$ with an uniaxial out-of-plane MAE of 0.775 meV/f.c. (6.0×10$^6$ erg/cm$^3$), originating from the high susceptibility of Fe atoms surrounding the interstitial N based on theoretical calculations [24]. More recently, it is reported that a large quantity of powder of α''-$Fe_8N$ can be produced at low temperatures by spark plasma sintering [25]. Apparently, carbonization, boronization, and nitrogenation reactions can induce a stable tetragonal distortion hence an enhanced MAE in α-Fe. Particularly, nitrogenation reaction tends to give rise to negative formation energies, meaning more iron-nitride compounds can be designed. Apart from permanent magnets, iron-nitride also has the potential to be applied in MCE applications. As reported in Ref. [26], the ε-$Fe_3N$ and γ'-$Fe_4N$ are potential good MCE materials due to their large magnetic deformation.

Motivated by the above, we have conducted a systematic high-throughput screening for Fe-N intermetallic compounds based on density functional theory (DFT) calculations, especially focusing on the rare-earth-free permanent magnet and giant magnetocaloric applications. In our recent research paper [27], we have identified 49 ferromagnetic (FM) compounds with thermodynamic (convex hull smaller than 0.075 eV/at.), mechanical, and dynamical stabilities, focusing on the hardness. In this follow-up paper, we design the rare-earth-free permanent magnet and magnetocaloric applications based on the 49 stable magnetic $Fe_xN_{1-x}$ ($0<x<1$) alloys.

## 2. Computational details

As mentioned in our recent paper [28], the structures of the Fe-N candidates are obtained by the evolutionary algorithm from the USPEX 10.5.0 code (Universal Structure Prediction: Evolutionary Xtallography) [29–31]. The selected criteria for a structure are set as a negative formation energy, a convex hull smaller than 0.075 eV/at, mechanical (elastic constants) and dynamical (phonon spectral are calculated based on density functional perturbation theory by using Phonopy code [32,33]) stabilities. All calculations are performed by Vienna ab initio simulation package (VASP) based on density functional theory (DFT) using the projector augmented wave method [34,35] in the present research paper. The valence states are selected as $2s^22p^3$ and $3d^74s^1$ for N and Fe atoms. The cutoff energy for the plane wave is set to 500 (450) eV, and the energy convergence criterion is set to $1\times10^{-7}$ ($1\times10^{-6}$) eV with a *k*-space density of 50 (30) Å$^{-1}$ for MAE (structure relaxation) calculations.

## 3. Results and discussion

### 3.1 MAE

As known, magneto-crystalline anisotropy energy plays a fundamental role in the application of magnetic materials. As to permanent magnets, MAE is the atomic origination of the macroscopic coercivity. The broken continuous symmetry leads to energy dependence on the orientation of the spin in magnetic materials, *i.e.* MAE is expressed by

$$K_{\hat{n}_1-\hat{n}_2} = E_{\hat{n}_1} - E_{\hat{n}_2}, \quad (1)$$

where $E_{\hat{n}_{1/2}}$ denotes the energy for magnetization along the direction of $\hat{n}_{1/2}$. For a material with an arbitrary structure, the selected orientations are [001], [010], and [100], leading to three possible MAE values, namely $K_{100-001}$, $K_{001-010}$, and $K_{100-010}$. The distribution of maximum absolute MAE ($|K_{max}|$) as a function of the square of saturation magnetization is shown in Fig. 1 for the Fe-N compounds, comparing with the experimentally realized permanent magnets [1,8,36]. Overall, the 49 FM compounds have filled in the empty region of the application spectral between MAE and MS. Notably, we have summarized 15 candidates with a sizable MAE (at least one MAE larger than 0.4 MJ/m$^3$) in Table A1, which are potential rare-earth-free gapped permanent magnets filling in the gap of application spectral between high-performance permanent magnets (*i.e.* FePd and $CoPt_3$) with underlying heavy element and widely used transition-metal based permanent magnet (AlNiCo and ferrites). In this point of view, the Fe-N compounds can spread the application range of permanent magnets.

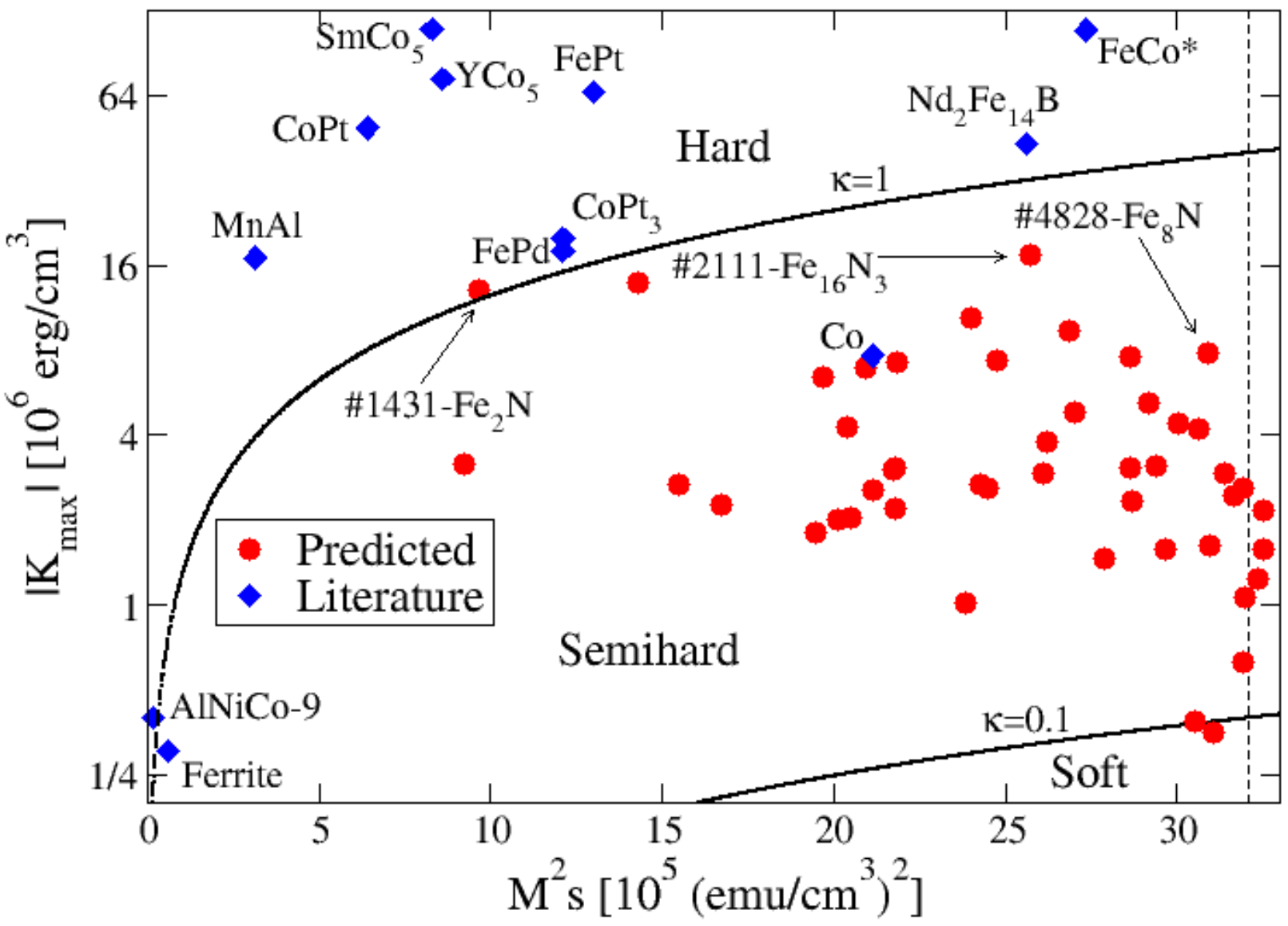

**Fig. 1.** The application spectrum of maximum absolute magneto-crystalline anisotropy ($K_{max}$) *vs.* saturation magnetization ($M_S$) for ferromagnetic $Fe_xN_{1-x}$ ($0<x<1$) alloys. The filled blue diamond and red circle symbols represent the data sets of well-known permanent magnets and the predicted $Fe_xN_{1-x}$ ($0<x<1$) alloys, respectively. The solid black lines correspond to the magnetic hardness of the compounds given by $\kappa = \sqrt{K_1/(\mu_0 M_S)^2}$, where $\kappa>1$, $0.1<\kappa<1$, and $\kappa<0.1$ represent the hard, semi-hard, and soft regions. The dashed vertical line indicates that the magnitude of magnetization is equal to that of pure α-iron.

For validation, we have compared $M_S$ and MAE with the associated data from references. For instance, α''-$Fe_8N$ (#4828) is also detected in our calculations with an out-of-plane MAE of 0.786 MJ/m$^3$ (0.786 meV/f.c.) and $M_S$ of 1758.35 emu/cm$^3$ in our calculations. Such values can be well comparable with the experimentally observed saturation magnetization (about 1750 emu/cm$^3$) and calculated MAE (0.775 meV/f.c. or 0.6 MJ/m$^3$) in Ref. [24]. For ε-$Fe_3N$ (#1908), the $M_S$ is 1419.12 emu/cm$^3$ (2.060 $\mu_B$/Fe) with an in-plane MAE of 0.202 MJ/m$^3$, where the density is 7.42 g/cm$^3$ based on Material Project [37]. The saturation magnetization can be compared with the experimental result of the ε-$Fe_3N$ nanoparticles (1.5 $\mu_B$/Fe, 1143 emu/cm$^3$) [38], while the MAE is roughly comparable with the result of the thin film ε-$Fe_{3.02}N$ (0.0326 MJ/m$^3$, based on the coercivity $H_C$ =207 Oe at room temperature) [39]. The comparable data indicates the reliability of our calculations.

Among all the Fe-N alloys (including the newly predicted cases as well as α''-$Fe_8N$ and ε-$Fe_3N$), it is found that the triclinic $Fe_{16}N_3$ (#2111) compound has the largest out-of-plane MAE of 1.751 MJ/m$^3$ (0.139 meV/Fe) with also a remarkable high $M_S$ of 1603.307 emu/cm$^3$ (1.850 $\mu_B$/Fe). It is noticed that such MAE value is more than twice that of that of α''-$Fe_8N$, and can be comparable with that of the well-known permanent magnets $L1_0$-type MnAl ($K_u$=1.525 MJ/m$^3$) [40] and FePd ($K_u$=1.8 MJ/m$^3$) [41], while the $M_S$ of $Fe_{16}N_3$ can even be comparable with that of the high-performance permanent magnet $Nd_2Fe_{14}B$ ($M_S$=1600 emu/cm$^3$) [1,42]. Moreover, we also find three other candidates with a MAE larger than 1 MJ/m$^3$ but an in-plane magnetization, *i.e.* the hexagonal #6365-$Fe_7N_3$ (MAE=1.384 MJ/m$^3$, $M_S$=1195.632 emu/cm$^3$), orthorhombic #1431-$Fe_2N$ (MAE=1.300

MJ/m$^3$, $M_S$=983.927 emu/cm$^3$), and orthorhombic #5076-$Fe_4N$ (MAE=1.047 MJ/m$^3$, $M_S$=1548.482 emu/cm$^3$). Particularly, #1431-$Fe_2N$ is a hard magnet because the dimensionless figure of merit $\kappa >$ 1 ($\kappa = \sqrt{K_1/(\mu_0 M_S)^2}$) [43], which can be applied in various situations since the MAE is retained regardless of the manufacturing shape [2,36]. In addition, the other 14 predicted potential rare-earth-free gapped permanent magnets (in Table A1) are all semi-hard magnets due to 0.1<$\kappa$<1, which are malleable and can be machined with standard metal-working tools [36,44].

Although with weak MAEs, the compounds #7124-$Fe_{20}N$ (triclinic), #2879-$Fe_{16}N$ (orthorhombic), #2518-$Fe_{19}N$ (triclinic), #4467-$Fe_{15}N$ (triclinic), #3110-$Fe_{14}N$ (triclinic), and #1966-$Fe_{12}N$ (monoclinic) respectively have significantly large saturation magnetizations of 1803.574 emu/cm$^3$ (MAE=0.157 MJ/m$^3$), 1802.684 emu/cm$^3$ (MAE=0.216 MJ/m$^3$), 1799.372 emu/cm$^3$ (MAE=0.123 MJ/m$^3$), 1788.767 emu/cm$^3$ (MAE=0.106 MJ/m$^3$), 1787.317 emu/cm$^3$ (MAE=0.063 MJ/m$^3$), and 1786.716 emu/cm$^3$ (MAE=0.261 MJ/m$^3$), which are larger than that of pure Fe (1784.761 emu/cm$^3$, based on our DFT calculations). The distribution of $M_S$ and the content ratio ($x$) of Fe is shown in Fig. 2 for $Fe_xN_{1-x}$ (0<$x$<1) intermetallic compounds. Obviously, the saturation magnetization increases approximately linearly with the increasing of the content ratio of Fe. On the other hand, the different chemical environment can induce different magnitude of saturation magnetization for compounds with the same content ratio of Fe.

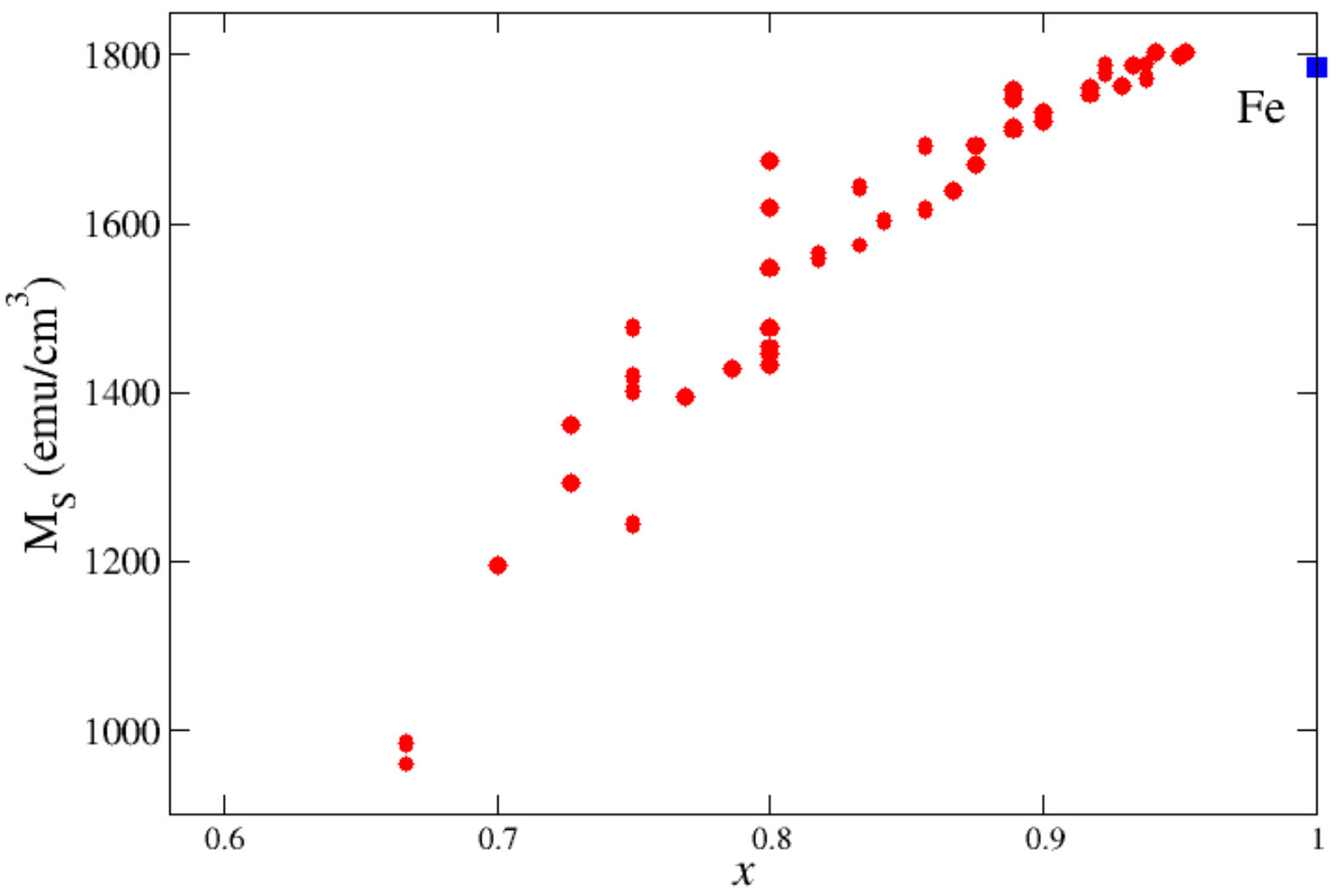

**Fig. 2.** The distribution of saturation magnetization ($M_S$) with respect to the content ratio $x$ in $Fe_xN_{1-x}$ (0<$x$<1) compounds. The red circle stands for the data set of the predicted compounds, while the blue square is the data of the pure α-iron.

### 3.2 MCE

Turning now to MCE applications, the potential fine MCE candidates can be easily screened by DFT calculations based on magnetic deformation ($\Sigma_M$) proxy, which is strongly associated with the essential parameter of entropy change $\Delta S_M$ for magneto-structural phase transition in MCE application [26]. The magnetic deformation is defined as a percentage indicator in terms of the degree of the Lagrangian finite strain tensor of lattice deformation ($\boldsymbol{\eta}$)

$$\Sigma_{\mathrm{M}} = \frac{1}{3}\sqrt{\eta_1^2 + \eta_2^2 + \eta_3^2} \times 100, \tag{2}$$

$$\boldsymbol{\eta} = \frac{1}{2}(\boldsymbol{P}^{\mathrm{T}}\boldsymbol{P} - \boldsymbol{I}). \tag{3}$$

Here, $\boldsymbol{P} = \boldsymbol{A}_{\mathrm{NM}}^{-1} \cdot \boldsymbol{A}_{\mathrm{M}}$, where $A_{\mathrm{NM}}$ (or $A_{\mathrm{M}}$) denotes the lattice constant in non-magnetic/magnetic structure. It is demonstrated that only with a magnetic deformation $\Sigma_M > 1.5\%$ can a compound be a potential MCE material [26].

We have found 40 newly potential MCE candidates ($\Sigma_M > 1.5\%$) among Fe-N alloys, where the relationship between saturation magnetization and magnetic deformation is shown in Fig. 3, also listing in Table A2. The magnetic deformation proxies for ε-$Fe_3N$ (#1908) and γ'-$Fe_4N$ (#2105) are respectively 1.93% and 1.92%, which are in good agreement with the results (1.88% and 1.93%) reported by Bocarsly *et al.* in Ref. [26]. Apparently, the correlation between $M_S$ and $\Sigma_M$ can be roughly regarded as positive in Fe-N alloys, which is similar to the relationship in MAB phase compounds [23]. Overall, we have found 41 compounds with a magnetic deformation $\Sigma_M > 1.5\%$, where 17 compounds even have the magnetic deformation larger than 7%. Particularly, it is found that the magnetic the deformation for the triclinic $Fe_7N$ (#2952) is surprisingly as large as 9.36%. It is noted the magnetic deformation for pure α-Fe is 1.57%. Moreover, it is found that the most promising permanent magnets (MAE>1 MJ/m$^3$) #2111-$Fe_{16}N_3$, #6365-$Fe_7N_3$, #1431-$Fe_2N$, and #5076-$Fe_4N$ are also potential MCE candidates, where the magnetic deformations are respectively 6.65%, 2.24%, 1.78%, and 4.69%. In this regard, such four compounds are promising candidates in both permanent magnet and giant magnetocaloric effect applications. All in all, we suspect that the Fe-N alloy system is a good playground to realize giant magnetocaloric effect applications.

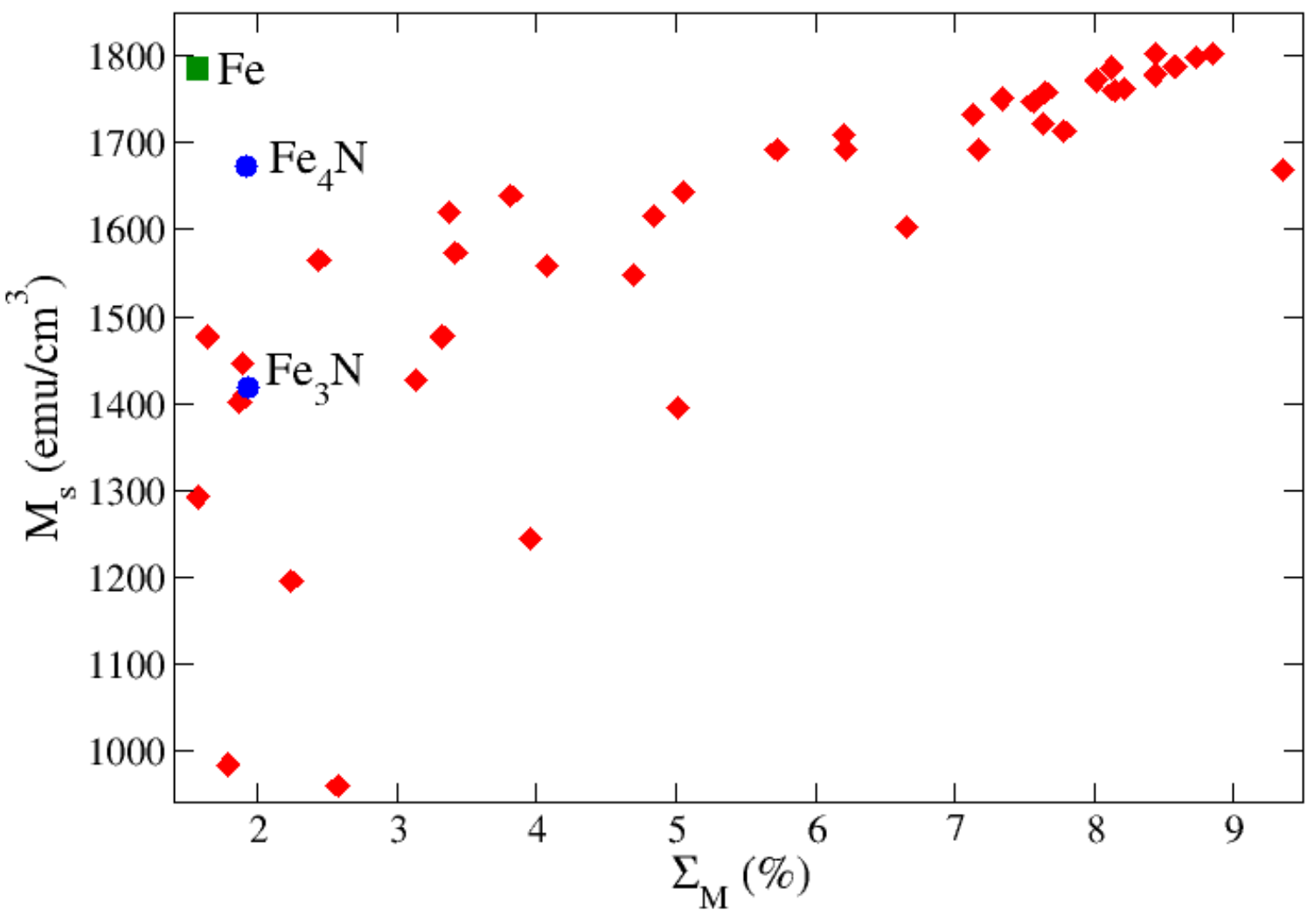

**Fig. 3.** The relationship between saturation magnetization ($M_S$) and magnetic deformation ($\Sigma_M$) for the 40 potential giant magnetocaloric effect candidates ($\Sigma_{\mathrm{M}} > 1.5\%$), displayed by the red diamonds. The green square represents the data set of the pure α-Fe (based on our calculations), while the blue circles are the data points of ε-$Fe_3N$ and $\gamma$'-$Fe_4N$.

## 4. Conclusions

In summary, we have carried out a systematic high-throughput DFT screening for permanent magnets and giant magnetocaloric effect candidates in $Fe_xN_{1-x}$ ($0<x<1$) intermetallic compounds. In total, 49 ferromagnetic compounds are identified in the Fe-N system. It is found that 15 compounds are potential permanent magnets with a magneto-crystalline anisotropy energy larger than 0.4 $MJ/m^3$, where $Fe_2N$ is a hard magnet while the other 14 candidates are semi-hard magnets. Such predicted permanent magnets can be applied as rare-earth-free gapped permanent magnets, spreading the application spectrum. Particularly, there are four promising permanent magnet candidates with a MAE more than 1 $MJ/m^3$. The saturation magnetization and the content ratio $x$ of Fe behaves like an approximate linear relationship in $Fe_xN_{1-x}$ ($0<x<1$). Based on the magnetic deformation proxy, 40 newly potential giant magnetocaloric effect candidates are discovered among $Fe_xN_{1-x}$ ($0<x<1$) compounds, where the magnetic deformation of $Fe_7N$ is significantly large as 9.36%. The $Fe_xN_{1-x}$ compound is a fine playground to design permanent magnets and giant magnetocaloric effect applications.

### CRediT authorship contribution statement

**Qiang Gao:** Conceptualization, Data curation, Formal analysis, Investigation, Methodology, Validation, Visualization, Writing – original draft, Writing – review and editing, **Ergen Bao:** Data curation, Formal analysis, Investigation, Validation, **Ijaz Shahid:** Data curation, Formal analysis, **Hui Ma:** Conceptualization, Formal analysis, Investigation, Methodology, Project administration, Writing – review and editing, **Xing-Qiu Chen:** Funding acquisition, Project administration, Software, Supervision, Writing – review and editing.

### Declaration of Competing Interest

The authors declare that they have no known competing financial interests or personal relationships that could have appeared to influence the work reported in this paper.

### Data availability

The structural 'POSCAR' for each compound is listed in the supplementary material with job id in the first line. The data that support the findings of this study are available from the corresponding author upon request.

### Acknowledgements

This work was supported by National Science and Technology Major Project of China (No. J2019-VI-0019-0134), Liaoning Provincial Natural Science Foundation Project of China (No. 2023-MS-017) and National Natural Science Foundation of China (No. 52188101).

**Appendix A.** Summaries of predicted permanent magnets and potential giant magnetocaloric effect candidates in $Fe_xN_{1-x}$ Compounds.

**Table A1:** The basic information for the predicted most promising permanent magnets (MAE larger than 0.4 MJ/m$^3$) in $Fe_xN_{1-x}$ ($0<x<1$) compounds. ID represents the job ID given by USPEX code. Com., ST, and SG denote the abbreviations of compound formula, structure type, and space group. In the ST column, TRI, HEX, ORT, MON, TET, and CUB mark the triclinic, hexagonal, orthorhombic, monoclinic, tetragonal and cubic crystal structure, respectively. $\Delta E_f$ and $\Delta E_H$ represent the formation and convex hull both in the unit of eV/at. MAE is the magneto-crystalline anisotropy energy in units of MJ/m$^3$ and meV/Fe. $M_S$ is the saturation magnetization in units of emu/cm$^3$ and $\mu_B$/Fe. The data (in bold font) of #4828-$Fe_8N$ (α''-$Fe_8N$) and #1908-$Fe_3N$ (ε-$Fe_3N$) are also listed for comparison with references. The formation and convex hull can be checked in our recent paper [27].

| ID | Com. | ST | SG | $\Delta E_f$ | $\Delta E_H$ | MAE | Easy-axis | $M_S$ |
|---|---|---|---|---|---|---|---|---|
| 2111 | $Fe_{16}N_3$ | TRI | $P\bar{1}$ | -0.0004 | 0.0464 | 1.7509 (0.1389) | *z* | 1603.307 (1.850) |
| 6365 | $Fe_7N_3$ | HEX | $P6_3$ | -0.0452 | 0.041 | 1.3840 (0.1193) | *y* | 1195.632 (1.247) |
| 1431 | $Fe_2N$ | ORT | *Pbcn* | -0.0431 | 0.0511 | 1.2995 (0.1138) | *x* | 983.9266 (0.992) |
| 5076 | $Fe_4N$ | ORT | *Cmcm* | -0.0073 | 0.0519 | 1.0470 (0.0852) | *x* | 1548.482 (1.742) |
| 2958 | $Fe_{13}N_2$ | MON | *C2* | -0.0002 | 0.0393 | 0.9369 (0.0736) | *y* | 1639.209 (1.928) |
| 2421 | $Fe_7N$ | MON | *C2/m* | -0.0122 | 0.0248 | 0.7536 (0.0590) | *x* | 1691.75 (2.003) |
| 1609 | $Fe_5N$ | MON | *C2* | -0.0077 | 0.0416 | 0.7376 (0.0591) | *y* | 1573.671 (1.815) |
| 840 | $Fe_3N$ | MON | *C2/m* | -0.0176 | 0.0564 | 0.7283 (0.0619) | *z* | 1477.306 (1.628) |
| 2018 | $Fe_4N$ | ORT | *Pmna* | -0.0313 | 0.0279 | 0.6929 (0.0559) | *z* | 1445.983 (1.611) |
| 1546 | $Fe_3N$ | ORT | *Immm* | -0.0088 | 0.0653 | 0.6385 (0.0535) | *y* | 1402.264 (1.523) |
| 8109 | $Fe_8N$ | TRI | *P*1 | -0.0166 | 0.0163 | 0.5201 (0.0403) | *z* | 1708.572 (2.034) |
| 617 | $Fe_5N$ | MON | *C2/m* | -0.0069 | 0.0424 | 0.4829 (0.0388) | *y* | 1643.37 (1.903) |
| 1266 | $Fe_9N$ | MON | *C2/m* | -0.0157 | 0.0139 | 0.4432 (0.0341) | *z* | 1732.425 (2.070) |
| 7844 | $Fe_{11}N_3$ | TRI | $P\bar{1}$ | -0.0003 | 0.0632 | 0.4246 (0.0343) | *x* | 1427.379 (1.647) |
| 3913 | $Fe_{11}N$ | MON | *C2/m* | -0.011 | 0.0137 | 0.4234 (0.0322) | *z* | 1750.894 (2.106) |
| 4828 | $Fe_8N$ | TET | *I4/mmm* | -0.0289 | 0.0040 | 0.7856 (0.0614) | *z* | 1758.348 (2.111) |
| 1908 | $Fe_3N$ | HEX | $P6_322$ | -0.0740 | 0.0000 | 0.2021 (0.0337) | *y* | 1419.120 (2.060) |

**Table A2**: The basic information for the predicted potential giant magnetocaloric effect candidates ($\Sigma_M > 1.5\%$) in $Fe_xN_{1-x}$ (0<$x$<1) compounds. The data (in green) of pure α-Fe together with that of (in blue) $Fe_3N$ (#1908) and $Fe_4N$ (#2105) are also listed for comparison.

| ID | Com. | ST | SG | $\Delta E_f$ | $\Delta E_H$ | $\Sigma_M$ | $M_S$ |
|---|---|---|---|---|---|---|---|
| 2952 | $Fe_7N$ | TRI | *P*1 | -0.0017 | 0.0353 | 9.36 | 1669.120 |
| 7124 | $Fe_{20}N$ | TRI | $P\bar{1}$ | -0.0042 | 0.0099 | 8.85 | 1803.574 |
| 2518 | $Fe_{19}N$ | TRI | $P\bar{1}$ | -0.0066 | 0.0082 | 8.74 | 1799.372 |
| 3110 | $Fe_{14}N$ | TRI | $P\bar{1}$ | -0.0129 | 0.0069 | 8.58 | 1787.317 |
| 4467 | $Fe_{15}N$ | TRI | $P\bar{1}$ | -0.0095 | 0.0090 | 8.57 | 1788.767 |
| 2879 | $Fe_{16}N$ | ORT | *Fmmm* | -0.0066 | 0.0108 | 8.44 | 1802.684 |
| 2466 | $Fe_{12}N$ | TRI | $P\bar{1}$ | -0.0192 | 0.0035 | 8.44 | 1778.582 |
| 5565 | $Fe_{13}N$ | TRI | $P\bar{1}$ | -0.0038 | 0.0173 | 8.22 | 1762.738 |
| 2226 | $Fe_{11}N$ | TRI | $P\bar{1}$ | -0.0147 | 0.0100 | 8.14 | 1760.069 |
| 1966 | $Fe_{12}N$ | MON | *C2/m* | -0.0120 | 0.0108 | 8.13 | 1786.716 |
| 3725 | $Fe_{15}N$ | MON | *C2/m* | -0.0066 | 0.0119 | 8.02 | 1772.071 |
| 510 | $Fe_8N$ | TRI | $P\bar{1}$ | -0.0031 | 0.0298 | 7.79 | 1714.373 |
| 3603 | $Fe_9N$ | TRI | $P\bar{1}$ | -0.0084 | 0.0212 | 7.63 | 1721.512 |
| 4010 | $Fe_8N$ | MON | *C2/m* | -0.0218 | 0.0111 | 7.56 | 1747.212 |
| 3913 | $Fe_{11}N$ | MON | *C2/m* | -0.0110 | 0.0137 | 7.34 | 1750.894 |
| 5471 | $Fe_7N$ | TRI | $P\bar{1}$ | -0.0110 | 0.0260 | 7.17 | 1693.076 |
| 1266 | $Fe_9N$ | MON | *C2/m* | -0.0157 | 0.0139 | 7.13 | 1732.425 |
| 2111 | $Fe_{16}N_3$ | TRI | $P\bar{1}$ | -0.0004 | 0.0464 | 6.65 | 1603.307 |
| 2421 | $Fe_7N$ | MON | *C2/m* | -0.0122 | 0.0248 | 6.22 | 1691.750 |
| 8109 | $Fe_8N$ | TRI | *P1* | -0.0166 | 0.0163 | 6.20 | 1708.572 |
| 5266 | $Fe_6N$ | MON | *C2/m* | -0.0078 | 0.0345 | 5.72 | 1692.724 |
| 617 | $Fe_5N$ | MON | *C2/m* | -0.0069 | 0.0424 | 5.05 | 1643.370 |
| 6067 | $Fe_{10}N_3$ | TRI | *P*1 | -0.0080 | 0.0604 | 5.02 | 1395.788 |
| 4811 | $Fe_6N$ | TRI | *P*1 | -0.0144 | 0.0279 | 4.84 | 1615.881 |
| 5076 | $Fe_4N$ | ORT | *Cmcm* | -0.0073 | 0.0519 | 4.69 | 1548.482 |
| 1516 | $Fe_9N_2$ | MON | *C*2 | -0.0013 | 0.0526 | 4.07 | 1558.602 |
| 687 | $Fe_3N$ | MON | *Pm* | -0.0077 | 0.0663 | 3.96 | 1243.857 |
| 2958 | $Fe_{13}N_2$ | MON | *C*2 | -0.0002 | 0.0393 | 3.81 | 1639.209 |
| 1609 | $Fe_5N$ | MON | *C*2 | -0.0077 | 0.0416 | 3.42 | 1573.671 |
| 3594 | $Fe_4N$ | ORT | *Fmmm* | -0.0040 | 0.0552 | 3.37 | 1619.769 |
| 2458 | $Fe_4N$ | TRI | $P\bar{1}$ | -0.0161 | 0.0431 | 3.32 | 1476.892 |
| 7844 | $Fe_{11}N_3$ | TRI | $P\bar{1}$ | -0.0003 | 0.0632 | 3.13 | 1427.379 |
| 1080 | $Fe_2N$ | ORT | *Pnnm* | -0.0283 | 0.0659 | 2.57 | 959.379 |
| 8255 | $Fe_9N_2$ | MON | *C2/m* | -0.0076 | 0.0462 | 2.44 | 1564.765 |
| 6365 | $Fe_7N_3$ | HEX | $P6_3$ | -0.0452 | 0.0410 | 2.24 | 1195.632 |
| 2018 | $Fe_4N$ | ORT | *Pmna* | -0.0313 | 0.0279 | 1.89 | 1445.983 |
| 1546 | $Fe_3N$ | ORT | *Immm* | -0.0088 | 0.0653 | 1.87 | 1402.264 |
| 1431 | $Fe_2N$ | ORT | *Pbcn* | -0.0431 | 0.0511 | 1.78 | 983.927 |
| 3566 | $Fe_4N$ | ORT | *Pbcn* | -0.0253 | 0.0339 | 1.64 | 1476.661 |

**Table A2** (*Continued.*)

| | | | | | | | |
|---|---|---|---|---|---|---|---|
| 840 | $Fe_3N$ | MON | *C*2/*m* | -0.0176 | 0.0564 | 1.64 | 1477.306 |
| 1617 | $Fe_8N_3$ | TRI | *P*1 | -0.0192 | 0.0604 | 1.57 | 1292.213 |
| 1 | Fe | CUB | $Im\bar{3}m$ | 0 | 0 | 1.57 | 1784.761 |
| 1908 | $Fe_3N$ | HEX | $P6_322$ | -0.0740 | 0 | 1.93 | 1419.120 |
| 2105 | $Fe_4N$ | CUB | $Pm\bar{3}m$ | -0.0396 | 0.0196 | 1.92 | 1674.060 |